\documentclass[runningheads]{llncs}
\usepackage{array}
\usepackage{booktabs}
\usepackage{enumitem}
\usepackage{hyperref}
\usepackage[T1]{fontenc}
\usepackage{float}
\usepackage[section]{placeins}
\usepackage[ruled,linesnumbered,noend,noline]{algorithm2e}
\usepackage{geometry}
\usepackage{graphicx}
\usepackage{algpseudocode}
\usepackage{underscore}
\usepackage{amsmath}
\usepackage{authblk}
\usepackage{subcaption}
\usepackage[perpage]{footmisc}
\usepackage[table,xcdraw]{xcolor}
\usepackage[backend=biber,
style=numeric, 
citestyle=numeric-comp,
sorting=none, 
isbn=false, 
url=false,
eprint=false,
date=year, 
mincitenames=3,
maxcitenames=3, 
minbibnames=3, 
maxbibnames=3,
uniquename=false,
uniquelist=false,
giveninits=true]{biblatex}
\AtEveryBibitem{\clearfield{doi}}
\usepackage{bbding}
\newcommand{\myorcidID}[1]{\href{https://orcid.org/#1}{\includegraphics[width=8pt]{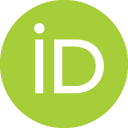}}}
\definecolor{customblue}{RGB}{135, 179, 224}
\begin{document}
	\title{UELLM: A Unified and Efficient Approach  for Large Language Model Inference Serving}
	\titlerunning{UELLM: A Unified and Efficient Approach  for LLM Inference Serving}
	%
	\author{Yiyuan He\inst{1,2}\myorcidID{0009-0003-2128-2852}\and
		Minxian Xu\inst{1(}\Envelope\inst{)}\myorcidID{0000-0002-0046-5153}\and
		Jingfeng Wu\inst{1} \myorcidID{0009-0008-5766-7873} \and
		Wanyi Zheng\inst{3}\myorcidID{0009-0004-8606-8387}\and
		Kejiang Ye\inst{1}\myorcidID{0000-0001-8985-2792}\and
		Chengzhong Xu\inst{4}\myorcidID{0000-0001-9480-0356}}
	
	\authorrunning{Y. He et al.}

	\institute{
		Shenzhen Institute of Advanced Technology, Chinese Academy of Sciences, Shenzhen, China \and
		Southern University of Science and Technology, Shenzhen, China\\ 
		\and 
		Shenzhen University of Advanced Technology, Shenzhen, China \\
		\email{\{yy.he2, mx.xu, jf.wu2, wy.zheng, kj.ye\}@siat.ac.cn}
		\and
		State Key Lab of IoTSC, University of Macau, Macau, China
		\email{czxu@um.edu.mo}
		}
	
	
	\maketitle  
        \vspace{-0.6cm}
	\begin{abstract}
		In the context of Machine Learning as a Service (MLaaS) clouds, the extensive use of Large Language Models (LLMs) often requires efficient management of significant query loads. When providing real-time inference services, several challenges arise. Firstly, increasing the number of GPUs may lead to a decrease in inference speed due to heightened communication overhead, while an inadequate number of GPUs can lead to out-of-memory errors. Secondly, different deployment strategies need to be evaluated to guarantee optimal utilization and minimal inference latency. Lastly, inefficient orchestration of inference queries can easily lead to significant Service Level Objective (SLO) violations.
		To address these challenges, we propose a \underline{U}nified and \underline{E}fficient approach for \underline{L}arge \underline{L}anguage \underline{M}odel  inference serving (UELLM), which consists of three main components: 1) $resource$ $profiler$, 2) $batch$ $scheduler$, and 3) $LLM$ $deployer$. The $resource$ $profiler$ characterizes resource usage of inference queries by predicting resource demands based on a fine-tuned LLM. The $batch$ $scheduler$ effectively batches the queries profiled by the $resource$ $profiler$ based on batching algorithms, aiming to decrease inference delays while meeting SLO and efficient batch processing of inference queries. The $LLM$ $deployer$ can efficiently deploy LLMs by considering the current cluster hardware topology and LLM characteristics, enhancing resource utilization and reducing resource overhead.  
		UELLM minimizes resource overhead, reduces inference latency, and lowers SLO violation rates. Compared with state-of-the-art (SOTA) techniques, UELLM reduces the inference latency by $72.3\%$ to $90.3\%$, enhances GPU utilization by $1.2\times$ to $4.1\times$, and increases throughput by $1.92\times$ to $4.98\times$, it can also serve without violating the inference latency SLO.
		
		\vspace{-0.2cm}
	\end{abstract}
	\vspace{-1cm}
	\section{Introduction}
	\vspace{-0.3cm}
	The rapid development of deep learning has driven the emergence of large models, and technology companies are increasingly building large MLaaS clouds for model training and inference services.  Due to the massive number of inference requests in MLaaS cloud services (e.g., trillions daily on Facebook, billions monthly on OpenAI \cite{morphling}), most resources and costs are dedicated to inference services (e.g., up to 90\% in AWS, 70\% in Meta AI \cite{metaai}).  Moreover, inference services are often part of user-facing applications, thus they have strict latency requirements \cite{clipper}. Additionally, different inference requests have different latency requirements. For example, 98\% of inference services need to be served within 200 milliseconds, while recommendation services require responses in less than 100 milliseconds \cite{mark}. Although the serving time for LLM can be extended, reducing LLM inference latency remains a critical requirement to ensure a good user experience \cite{vllm}. 
	
	In a typical MLaaS workflow, developers use large datasets to train LLMs offline, then deploy multiple trained LLMs in the cloud to provide online inference services.
	Due to the increasing number of parameters in current  LLMs,
	distributed deployment is required \cite{jin2023s3}. This means that LLMs need to be deployed on more hardware accelerators. However, the more hardware accelerators deployed simultaneously, the greater the communication latency between different hardware accelerators, which can increase inference latency and the rate of SLO violations. Additionally, owing to the sequential execution characteristics of LLM inference, multiple hardware accelerators can only work serially during inference. This implies that as more hardware accelerators are deployed, the waiting time increases, which in turn reduces the utilization rate. 
	
	To improve inference efficiency, batch processing\footnotemark[1] \footnotetext[1]{Batch processing represents the processing of inference requests with batches, which is different from batching as discussed in this article. Here, batching refers to the effective scheduling among batches and the efficient combination of requests within each batch.}is often used. However, the majority of contemporary LLM architectures are based on Transformer \cite{vaswani2023attention}. The autoregressive characteristic of the self-attention layer in these architectures presents notable obstacles for model deployment and batching. Specifically, when generating a new token, the model needs to attend to the previous tokens of that token. To reduce iterations, this requires the model to retain all information of the previous tokens and store it in memory \cite{jin2023s3}.  For a particular inference computation, denote the batch size by $b$, the maximum length of input sequence by $s$, the maximum length of output sequence  by $n$, the hidden dimension of the transformer by $h$,  and the total number of transformer layers by $l$. The total number of bytes to store the Key-Value Cache (KV Cache) in peak is $4 \times blh(s + n)$ \cite{sheng2023flexgen}. Thus, the size of the KV Cache grows with the batch size and the maximum length of output sequences. This means that processing requests with similar output lengths in a batch can reduce redundant KV Cache and calculation load.
	In light of these challenges, we propose UELLM, which integrates batching requests and deploying LLMs efficiently. UELLM aims to maximize throughput, reduce inference latency, and lower SLO violation rates.
	UELLM primarily consists of three components: $resource$ $profiler$, $batch$ $scheduler$ and $LLM$ $deployer$. The $resource$ $profiler$ mainly uses a fine-tuned LLM to predict the output length of each request and obtain the SLO for each request to facilitate subsequent scheduling. The $batch$ $scheduler$ optimizes the combination of inference requests within a batch based on predicted output sequence length and schedules them according to the SLO, reducing SLO violation rates and inference latency. The $LLM$ $deployer$ strategically deploys LLMs based on the network topology of the current hardware system and the specific characteristics of the LLMs, enhancing GPU utilization and reducing inference latency. Finally, UELLM runs a backend monitoring program to detect erroneous predictions and adjust the allocated memory size to improve accuracy. By integrating these components, UELLM optimizes memory usage and scheduling during inference. This integration leads to a reduction in latency and SLO violation rates, while also enhancing resource utilization when deploying Transformer-based LLMs for text generation on GPUs.
	
	In summary, our key \textbf{contributions} are:
	\vspace{-0.3cm}
	\begin{itemize}
		\item We analyze the primary bottlenecks present in LLM inference services: 1) the challenge of efficiently batching diverse inference requests, and 2) the difficulty of effectively utilizing resources during LLM inference due to the extensive search space and diverse model structures.
		\item We propose UELLM, which can reasonably adjust batch combinations to reduce latency and resource overhead, improve throughput, and efficiently deploy LLMs, thereby increasing resource utilization, reducing latency, and minimizing resource overhead while meeting SLOs.
		\vspace{-0.05cm}
		\item We experimentally evaluate the effectiveness of UELLM on a realistic cluster. Compared with SOTA techniques, UELLM reduces the inference latency by $72.3\%$ to $90.3\%$, enhances GPU utilization by $1.2\times$ to $4.1\times$, and increases throughput by $1.92\times$ to $4.98\times$, it can serve without violating the inference latency SLO.
	\end{itemize}
	\vspace{-0.6cm}
	\section{Background and Motivation} 
	\vspace{-0.2cm}
	In this section, we will introduce the relevant background information regarding LLMs and the phenomena that have motivated our design.
	\vspace{-0.3cm}
	\subsection{Background: Generative LLM Consumes Large Amount of Memory}
	
	Currently, Transformer-based generative LLMs (such as ChatGPT \cite{gpt}, Llama \cite{metaai}, ChatGLM \cite{du2022glm}, etc.) are autoregressive. They share the common characteristic of predicting the most probable token based on past tokens. When not using batch processing, the model generates one token at a time, requiring $n$ iterations to generate a sequence of $n$ tokens. Each iteration involves an input token traversing the model, which consists of a stack of transformer layers, including an attention layer, two normalization layers, and two feed-forward layers. The self-attention layer uses information from past tokens to generate the next token. Through the self-attention mechanism, the model can weight each word in the input sequence, focusing on important contextual information. During each step of the generation process, the attention mechanism needs to compute queries, keys, and values. To avoid recalculating each time, the previously generated keys and values can be cached. This caching mechanism significantly reduces computation during the generation phase and increases generation speed. However, since the KV Cache sequentially stores information about previous tokens, it expands as the model generates more tokens.
	\vspace{-0.3cm}
	\subsection{Motivation: Deployment Configuration versus LLM Inference Performance }
	\vspace{-1.1cm}
	\begin{figure}[htbp]
		\centering
		\begin{subfigure}[b]{0.346\textwidth}
			\centering
			\includegraphics[width=\textwidth]{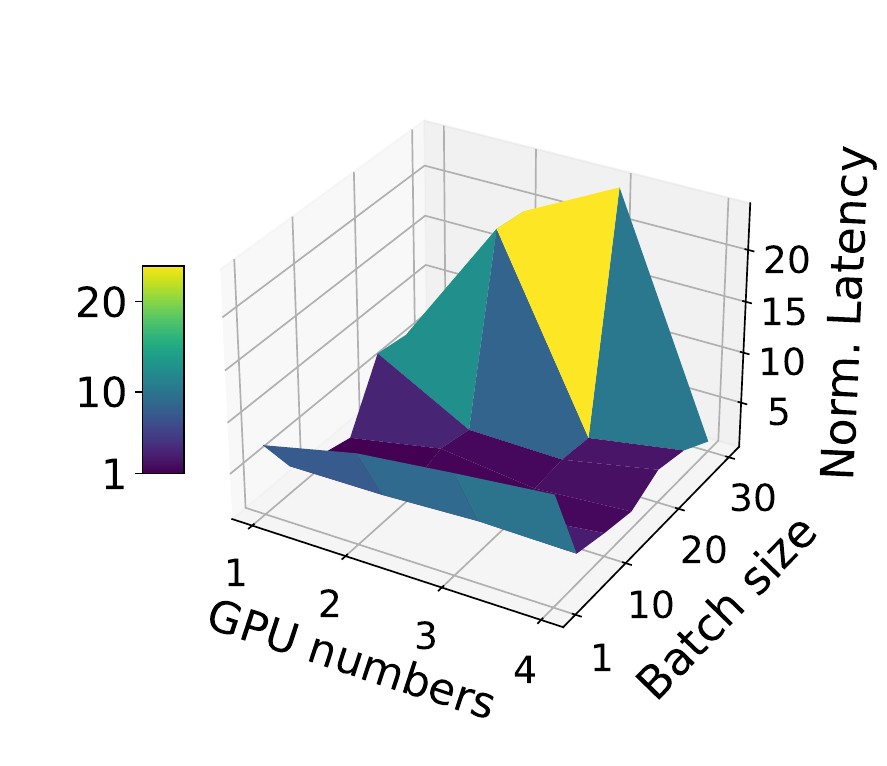}
			\caption{Latency.}
			\label{fig:image1}
		\end{subfigure}
		\hfill
		\begin{subfigure}[b]{0.315\textwidth}
			\centering
			\includegraphics[width=\textwidth]{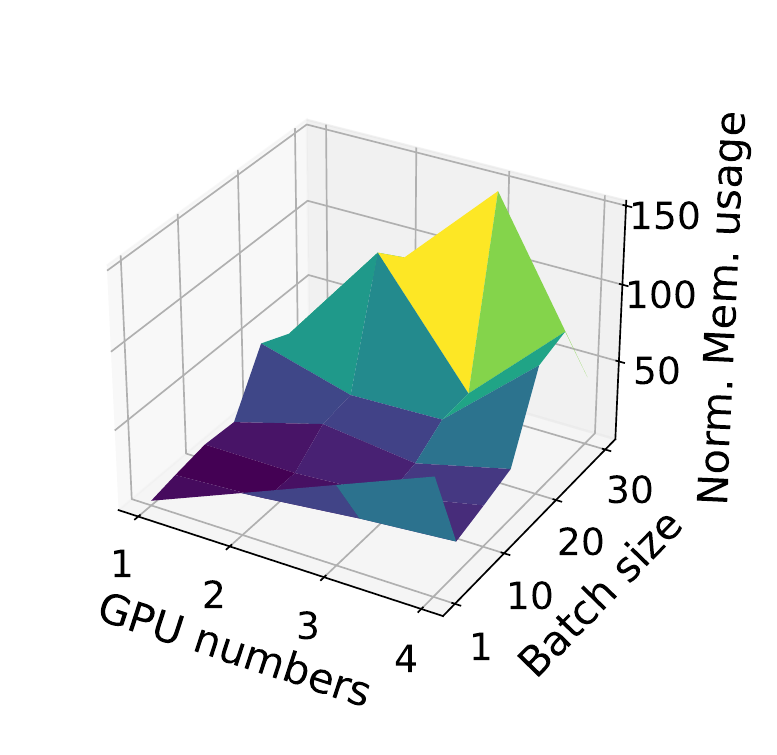}
			\caption{Memory usage.}
			\label{fig:image2}
		\end{subfigure}
		\hfill
		\begin{subfigure}[b]{0.313\textwidth}
			\centering
			\includegraphics[width=\textwidth]{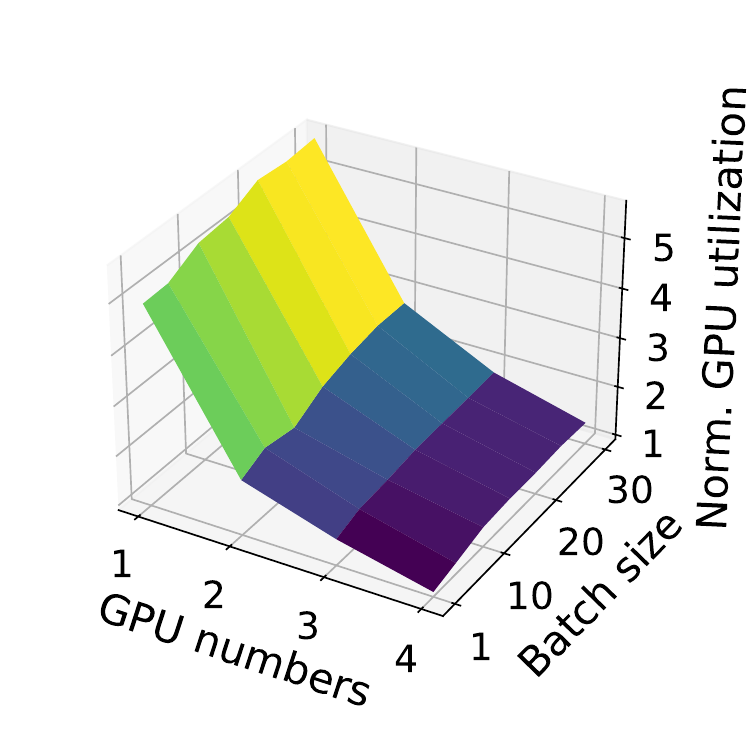}
			\caption{GPU Utilization.}
			\label{fig:image3}
		\end{subfigure}
		\caption{Normalized latency, memory usage, and GPU utilization under different configurations of GPU numbers and batch sizes. Each metric is normalized to its minimum value.}
		\label{fig:normal}
	\end{figure}
	\vspace{-0.6cm}
	\noindent\textbf{Observation \#1: Slight changes in deployment configurations can have a significant impact on LLM inference performance. }With the development of large model technologies, the size and parameters of LLMs have increased. To deploy LLMs for inference services, it is essential to allocate a significant amount of memory. Nevertheless, the growth in the size of LLMs far outpaces the development of hardware, making it difficult for a single hardware accelerator (e.g., GPU) to support a LLM. Consequently, when deploying LLMs for inference tasks, it is common practice to distribute the model across multiple hardware accelerators, thereby requiring a distributed deployment approach. To facilitate this, a device map is required to delineate the allocation of model layers to specific hardware accelerators. To ensure efficient subsequent inference, the device map must be finely tuned, as the inference performance of LLMs is highly sensitive to different deployment configurations. The process of a deployment configuration can be simplified as follows: \textbf{1)} Determining the number of GPUs, and \textbf{2)} Detailed adjustment of the device map to deploy each layer of LLM on the corresponding GPU. Determining the appropriate number of GPUs can be challenging, as merely increasing the number of GPUs may not always be advantageous, given that an excessive number of GPUs can escalate communication and synchronization costs among them. Fig. \ref{fig:normal}  shows that a reasonable number of GPUs can improve GPU utilization by 4$\times$ and reduce latency by 20$\times$ compared to a poor GPU configuration.  Even if the optimal number of GPUs is determined, fine-tuning the device map  is required. Table \ref{table:gpu} shows that a well-configured device map (last row in Table \ref{table:gpu}) has the potential to increase throughput significantly, doubling it compared to a poorly configured setup (increasing from 11.19 to 22.55 in terms of average throughput). Therefore, the inference performance of LLM is notably impacted by the model's deployment configurations.
	\vspace{-0.5cm}
	\begin{table}[htbp]
		\small
		\centering
		\begin{tabular}{>{\centering\arraybackslash}p{2.5cm}>{\centering\arraybackslash}p{2.5cm}>{\centering\arraybackslash}p{2.3cm}>{\centering\arraybackslash}p{2.3cm}>{\centering\arraybackslash}p{2.3cm}}
			
			\toprule
			\multicolumn{2}{c}{\cellcolor{customblue!20}\textbf{Device map}} & \multicolumn{3}{c}{\cellcolor{customblue!20}\textbf{Throughput (token/s)}} \\ 
			\cellcolor{customblue!10}\textbf{GPU\#0} & \cellcolor{customblue!10}\textbf{GPU\#1} & \cellcolor{customblue!10}\textbf{Average} & \cellcolor{customblue!10}\textbf{Maximum} & \cellcolor{customblue!10}\textbf{Minimum} \\ 
			\midrule
			\rowcolor{customblue!20}
			layer 0-15   & layer 16-32   & 11.19 & 11.58 & 10.84 \\ 
			\rowcolor{customblue!10}
			layer 0-19   & layer 20-32   & 13.09 & 13.48 & 12.61 \\ 
			\rowcolor{customblue!20}
			layer 0-23   & layer 24-32   & 14.85 & 15.45 & 14.09 \\ 
			\rowcolor{customblue!10}
			layer 0-27   & layer 28-32   & 17.23 & 18.00 & 16.16 \\ 
			\rowcolor{customblue!20}
			layer 0-31   & layer 32   & \textbf{22.55} & \textbf{23.07} & \textbf{22.11} \\ 
			\bottomrule
		\end{tabular}
		\vspace{0.3cm}
		\caption{throughput variations of ChatGLM2-6B on two GPUs (GPU\#0: Tesla V100 and GPU\#1: RTX 3090). It shows average, maximum, and minimum throughput for different device maps.}
		\label{table:gpu}
	\end{table}
	\vspace{-1.7cm}
	\subsection{Motivation: Batching versus LLM Inference Performance} \label{sec:batch}
	\vspace{-0.2cm}
	\textbf{Observation \#2: Batching multiple requests can reduce the SLO violation rate of LLM inference services under a large number of inference requests.}
	When faced with a large number of inference requests, batching multiple requests can reduce the SLO violation rate of LLM inference services. This is because different requests can share weights, allowing multiple tokens to be generated in a single iteration through batching, thereby increasing the token generation rate and reducing the SLO violation rate. However, batching requests for LLM services requires addressing two key issues. 1) Requests may arrive at different times. A simple batching strategy either makes earlier requests wait for later ones or delays incoming requests until earlier ones are processed, leading to significant queuing delays. 2) Requests may have vastly different input and output lengths. Current batching techniques pad the inputs and outputs of requests to balance their lengths \cite{jin2023s3}, but this strategy requires fine-tuning of the batched inference requests. Otherwise, it will lead to significant wastage of GPU computation and memory (see Section \ref{sec:batch}). Fig. \ref{fig:normal} shows the relationship between batch size, GPU numbers, latency, GPU utilization, and memory usage. It is evident that a well-configured combination of batch size and GPU numbers can reduce latency by nearly 20$\times$ (with the worst-case scenario involving offloading), improve GPU utilization by 5$\times$, and reduce memory usage by 150$\times$. Consequently, the default batch combination technique results in significant memory waste and low effective GPU utilization. 
	
	\vspace{-0.2cm}
	Given the background and our observations, they motivate us to design a comprehensive framework capable of efficiently deploying LLMs and effectively combining batches during inference, thereby improving system utilization and reducing inference latency.
	
	\vspace{-0.4cm}
	\section{Related Work}
	\vspace{-0.25cm}
	In this section, we will discuss the SOTA technologies focusing on model deployment resource allocation and inference request batching in LLM inference services, and their associated limitations.
    \vspace{-0.8cm}
	\subsection{Model Deployment}
	\vspace{-0.15cm}
	In the realm of model deployment resource allocation, Zhang et al. \cite{mark} proposed MArk, a service system designed for ML inference. MArk integrates IaaS and serverless computing to minimize costs while meeting SLOs. However, MArk primarily focuses on the effective deployment of small models, overlooking LLMs and the influence of device map on LLM deployment. In contrast, Wang et al. \cite{morphling} presented  Morphling, a rapid and nearly optimal auto-configuration framework for cloud-native model services. Morphling uses model-agnostic meta-learning to navigate large configuration spaces. Morphling quickly adapts the meta-model for new inference services by sampling a small number of configurations and utilizing it to find the best configuration. However, Morphling performs stress tests on each candidate configuration, resulting in additional computational burden and causing significant latency during LLM testing, rendering it unsuitable for scenarios with limited resources.
	
	Moreover, there are research studies that concentrate on the hardware aspect. Choi et al. \cite{choi} focused on the hardware layer and introduced a novel multi-model machine learning inference server scheduling framework. A crucial aspect of their proposal involves utilizing hardware support for the spatial partitioning of GPU resources. By implementing spatiotemporal sharing, they established a new GPU resource abstraction layer was created using configurable GPU resources. The scheduler assigns requests to virtual GPUs, known as gpulets, based on the optimal resource allocation. In order to reduce the expenses associated with cloud-based inference servers, the framework dynamically adjusts the number of GPUs required for a specific workload. Regrettably, Choi et al. solely focused on deploying small models (e.g., ResNet50, LeNet, VGG16) and did not consider the prevalent large models in the current landscape (e.g., ChatGLM \cite{du2022glm}, Llama \cite{metaai}).
	\vspace{-0.3cm}
	\subsection{Model Inference}
	In the context of batching inference requests, the predominant system in use is the Triton Inference Server\footnotemark[1]\footnotetext[1]{https://developer.nvidia.com/triton-inference-server}. It provides simple dynamic scheduling and batch processing features. As previously mentioned in Section \ref{sec:batch}, getting the output length of the query before scheduling  is crucial for reducing latency and enhancing resource utilization. Nevertheless, Triton Inference Server lacks a viable batch combination algorithm that takes into account SLOs and query output for scheduling. To address this problem, Jin et al. \cite{jin2023s3} presented $S^3$, which is a system-algorithm co-design framework that treats batch combination as a bin packing problem through sequence length prediction to maximize GPU utilization and achieve higher throughput. However, $S^3$ only considers predicted output length when scheduling requests, without accounting for SLOs and other metrics.

	Some studies focus on the inference services of traditional machine learning models. Ali et al. \cite{batch} proposed BATCH , a framework designed to enhance the efficiency of machine learning services on serverless platforms by addressing the absence of batch processing support. 
     The framework utilizes lightweight profiling techniques and analytical models to identify the optimal parameter configurations (i.e., memory size, batch size, and timeout) to improve system performance while meeting user-defined SLOs. However, BATCH uses exhaustive search methods for configuration, resulting in high time complexity and failing to meet real-time requirements. Clearly, due to the significant differences between LLMs and traditional machine learning models, these inference systems and studies are challenging to apply to more complex LLM inference systems. Wang et al. \cite{tabi} proposed Tabi , the first inference system addressing the resource overhead in increasingly large language models. It is a multi-stage inference engine driven by individual query feedback and leverages the latest ML advancements to optimize LLM inference latency for classification tasks. However, Tabi is only optimized for discriminative models (such as text recognition models) within the service framework, i.e., non-generative LLMs (GPT \cite{gpt}, GLM \cite{du2022glm}).  Gunasekaran et al. proposed Cocktail \cite{cocktail}, a cost-effective ensemble-based model service framework aimed at providing highly accurate predictions with minimal latency and reducing deployment costs in public cloud environments. Cocktail consists of two key components: 1) a dynamic model selection framework that reduces the number of models in the ensemble while meeting accuracy and latency requirements, 2) an adaptive resource management framework that employs a distributed proactive auto-scaling strategy to allocate resources efficiently to models. Although Cocktail focuses on model selection, it is more centered on ensemble learning rather than selecting individual models for inference. Seo et al. \cite{wonik} proposed an SLO-aware inference scheduler for heterogeneous processors on edge platforms. While it addresses the issue of selecting appropriate model services for inference tasks, the models discussed are only small traditional deep learning models and do not consider large models.  

	Our work, UELLM, is most similar to $S^3$ and Morphling. However, there are significant differences. Compared to $S^3$, firstly, $S^3$ recognizes the issues faced by batching but focuses solely on reducing inference latency and improving utilization. In contrast, UELLM considers factors like SLO in addition to those addressed by $S^3$ and proposes more diverse scheduling algorithms that support customization for different scheduling objectives. Secondly, $S^3$ does not consider efficient model deployment, whereas UELLM uses dynamic programming to optimize resource allocation during LLM deployment. Thirdly, while both $S^3$ and UELLM fine-tune LLMs for predicting output length, UELLM employs online learning, which is better suited for the real-time tasks faced by LLM inference services. Compared to Morphling, UELLM uses simpler and more varied deployment algorithms to efficiently deploy LLMs, whereas Morphling employs more complex and singular meta-learning to find the optimal deployment configuration. Additionally, Morphling generates multiple possible configurations and stress tests them simultaneously, greatly increasing system resource usage. In UELLM, only one configuration is generated at a time, and supports dynamic scaling.
	\vspace{-0.4cm}
	\section{System Design} 
\begin{figure}[ht]
		\centering
		\includegraphics[width=\textwidth ]{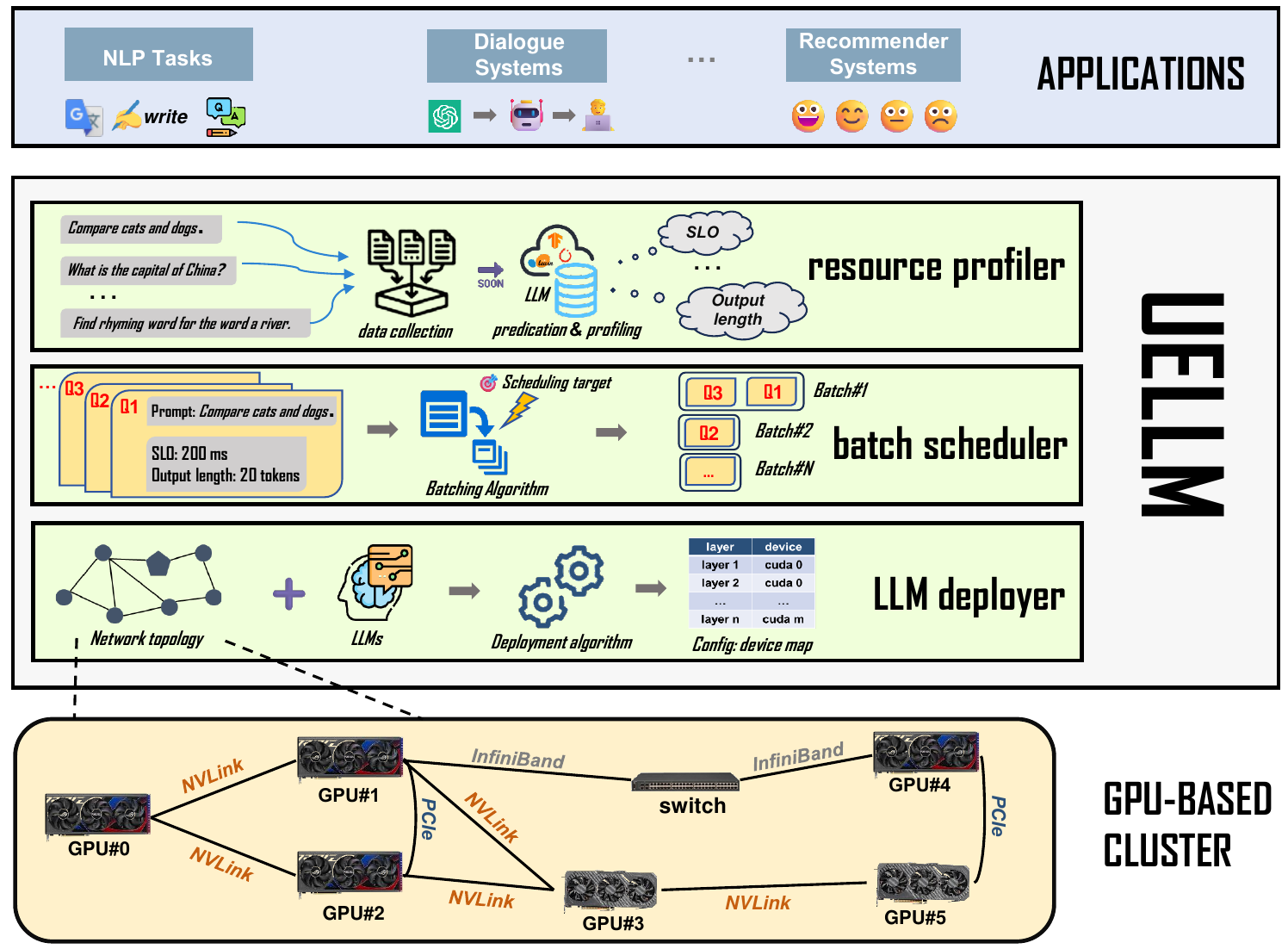} 
		\caption{The main framework and functional details of UELLM.}
		\label{fig:UELLM}
		\vspace{-0.3cm}
	\end{figure}

    \vspace{-0.15cm}
	In this section, we present the detailed design of UELLM, a unified LLM inference serving resource scheduling architecture comprising three main components: 1) the $resource$ $profiler$, 2) the $batch$ $scheduler$, and 3) the $LLM$ $deployer$. The primary framework is illustrated in Fig. \ref{fig:UELLM}.
	\vspace{-0.4cm}
	\subsection{Resource Profiler}
     \vspace{-0.15cm}
	
	Before scheduling, each request undergoes processing by the $resource$ $profiler$, which consists of three primary modules: data collection, output length prediction, and resource profiling. For predicting the length of inference requests, we draw inspiration from $S^3$ \cite{jin2023s3} and fine-tune the large model ChatGLM3-6B \cite{du2022glm} to categorize the output lengths. The model is fine-tuned on the representative Q\&A dataset  Alpaca\footnotemark[1], using questions as inputs and the token length of the answers as labels. Our observations indicate that the predictor accurately predicted the buckets with a precision of 99.51\%. Furthermore, we evaluate our predictor on established datasets such as the Google Natural-Question dataset\footnotemark[2] and the Alpaca GPT-4 dataset\footnotemark[1], where it consistently achieved an accuracy exceeding 80\%.
	\footnotetext[1]{https://github.com/tatsu-lab/stanford_alpaca}
	\footnotetext[2]{https://ai.google.com/research/NaturalQuestions}

         \vspace{-0.5cm}
    \vspace{-0.8cm}
	\subsection{Batch Scheduler} \label{batching}
\begin{figure}[htbp]
		\centering
		\includegraphics[width=\textwidth]{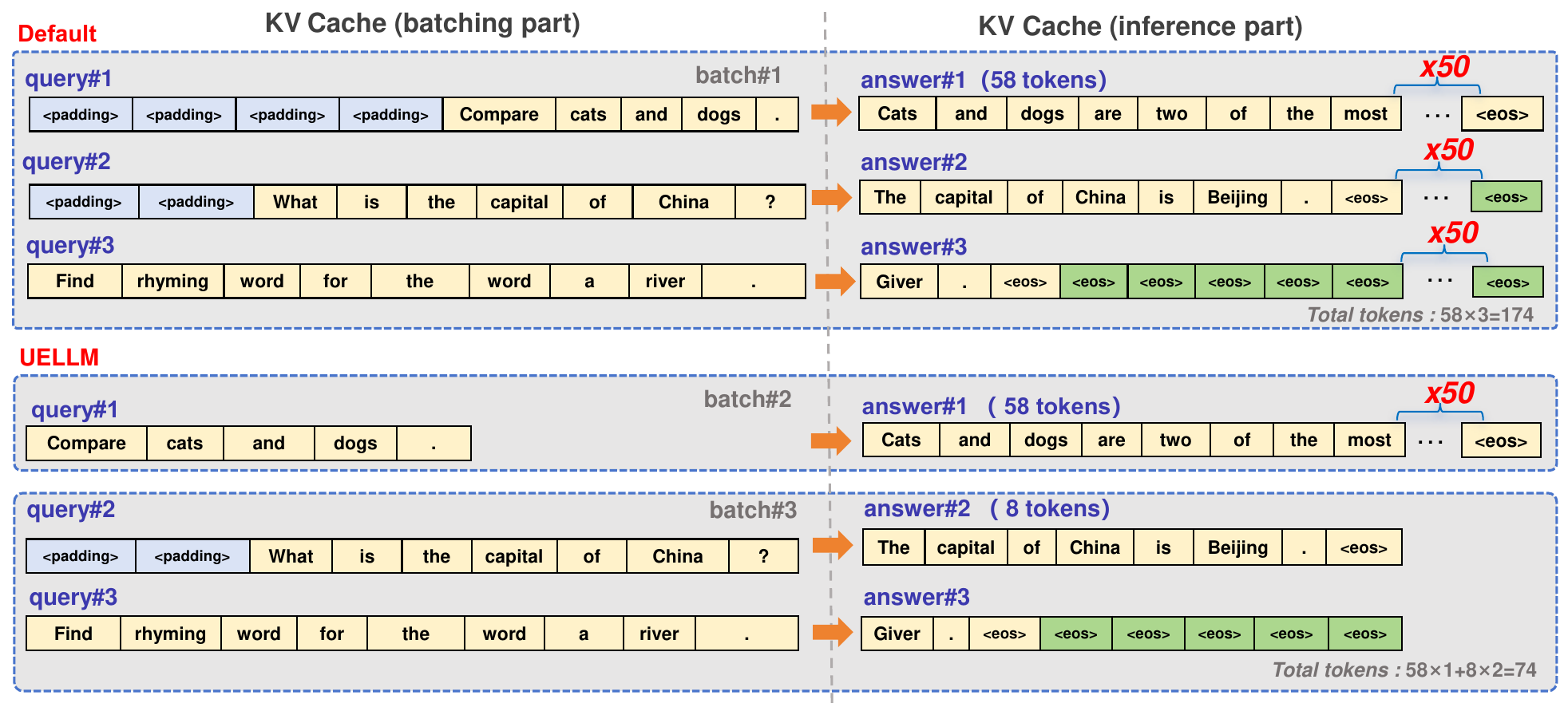} 
		\caption{Comparison between UELLM and the default batching algorithm. The comparison illustrates the utilization of the KV Cache in both the batching and inference stages for three queries. UELLM focuses on optimizing token usage, leading to a notable decrease in the overall number of tokens processed during inference in comparison to the default method.}
		\label{fig:1batch}
		\vspace{-0.6cm}
	\end{figure}

	\vspace{-0.1cm}
	Assume a $batch = \{q_1, q_2, ..., q_b\}$ contains $b$ queries, where each query $q_i$ indexed by $i$ has input and output lengths denoted as $Input_i$ and $Output_i$. The premise of model batching is to ensure that the lengths of all inputs are equal. Hence, prior to making inferences, each query will be padded to $\max_{i=1}^b (Input_i)$ to achieve uniform length, leading to increased memory usage as a result of padding. During the inference phase, the model needs to populate all outputs in the batch to $O = \max_{i=1}^b (Output_i)$. Thus the total number of generated tokens during inference is $b \times O$. In contrast to the substantial memory consumption associated with generating a large KV Cache, the memory occupied by batching itself is minimal \cite{vllm}. As Fig. \ref{fig:1batch} shows, if there are three queries: $query\#1$, $query\#2$, $query\#3$ to be scheduled,the default batching will batching these queries into a single batch, which will require 6 paddings and 174 tokens. Notably, for $query\#2$ and $query\#3$, numerous redundant tokens are generated. While UELLM will batching the three queries into two batches: $batch\#2$ and $batch\#3$, requiring only 2 paddings, generating 74 tokens and reducing the number of redundant tokens significantly. Since the number of tokens is roughly proportional to the computational and memory overhead, this reduction diminishes inference latency and memory usage. The efficacy of UELLM is validated in Section \ref{eva}.
     \vspace{-0.4cm}
	
	\vspace{-0.3cm}
	In UELLM, the $batch$ $scheduler$ plays a crucial role in batching inference requests following profiling into suitable batch  for subsequent LLM inference. The primary objective of the $batch$ $scheduler$ is to minimize latency, optimize memory usage, and prevent SLO violations during the batching process of these queries.
	To attain this objective, we introduced the SLO-ODBS algorithm \ref{slo-odbs}, which is based on the resource utilization characteristics during LLM inference, while considering both the inference latency and SLO requirements. Furthermore, to prevent excessive delays and overhead caused by the algorithm's execution, we have streamlined its design while ensuring its effectiveness. SLO-ODBS receives a sequence of inference requests from the resource profiler and outputs batches. The algorithm can be divided into three stages: \textbf{1)} the initialization phase (lines 1-4), \textbf{2)} the combining single batches based on output phase (lines 5-19), and \textbf{3)} the sorting all combined batches (lines 20-23).

	In stage \textbf{1)}, all requests are initially sorted in ascending order according to their SLO and a set of parameters is initialized.
	
	In stage \textbf{2)}, we first maintain the properties of the current batch to be combined, denoted as $batch_c$: the current maximum latency $L_{CM}$, the current maximum output length $O_{CM}$, and the current maximum composite metric $CM$. Therefore, the total latency $T_l$ can be defined as:
	\begin{equation}
		T_l=\sum_{i=1}^N\left(\left(SLO_i+L_{CM}\right)\times\left(|\mathrm{batch_c}|+1\right)\times L_1\right),
	\end{equation}
	and total output length $T_o$ can be denoted as:
	\begin{equation}
		T_o=\sum_{i=1}^N\left((Length_i+O_{CM})\times(|\mathrm{bacth_c}|+1)\times L_2\right).
	\end{equation}
	Among them, $N$ represents the number of queries in the $batch_c$ and $L1, L2$ represent the additional overhead due to parallel computing. Therefore, our optimization objective mainly considers two aspects: the total latency $T_l$ and the total output length $T_o$. We use weights $w_1$ and $w_2$ to balance the importance of these two factors:
	\begin{equation}
		\max_{w_1 , w_2}(w_1\times T_l+w_2\times T_o)  \leq Threshold.
	\end{equation}
	To optimize system performance, it is necessary to ensure that the total sum $Total$ of the batch does not exceed the threshold $T$ after adding each request:  $\mathrm{Total}\leq\mathrm{T}$. This approach allows the request sequence to be reassembled into $batch_1, batch_2, ..., batch_n$.

	In stage \textbf{3)}, we combine the batches obtained in stage \textbf{2)} into a ready list $batches$ = \{$batch_1$, $batch_2$, ..., $batch_n$\} for subsequent batch processing. 
	\vspace{-0.6cm}
	\begin{algorithm}[h]
		\caption{SLO and Output-Driven Dynamic Batch Scheduler (SLO-ODBS)} \label{slo-odbs}
		
		\KwIn{\textit{requests}: A list of requests after profiling}
		\KwOut{\textit{batches}: A list of batch}
		
		\SetKwFunction{SLOODBS}{SLO-ODBS}
		\SetKwProg{Fn}{Procedure}{:}{}
		\Fn{\SLOODBS{\textit{requests}}}{
			\textit{sorted\_requests} $\gets$ sort(\textit{requests}); \Comment{Sort by SLO in ascending order} \\
			\textit{batches, $batch_c$} $\gets \emptyset$;\\
			\textit{$L_{CM}$, $O_{CM}$, $CM$} $\gets$ 0\;
			
			\For{\textit{q} \textbf{in} \textit{sorted\_requests}}{
				$T_l$ $\gets$ (\textit{q.SLO} + \textit{$L_{CM}$}) $\times$ (\textit{len($batch_c$)} + 1) $\times$ $L1$\;
				$T_o$ $\gets$ (\textit{q.length} - \textit{$O_{CM}$}) $\times$ (\textit{len($batch_c$)} + 1) $\times$ $L2$\;
				\textit{Total} $\gets$ $w_1 \times \textit{Latency\_total} + w_2 \times \textit{Length\_total}$\;
				
				\If{\textit{$batch_c = \emptyset$} \textbf{or} \textit{Total} $\leq$ \textit{Threshold}}{
					\textit{$batch_c$.append(q.index)}\;
					\textit{$L_{CM}$} $\gets$ max(\textit{$L_{CM}$}, \textit{q.SLO})\;
					\textit{$O_{CM}$} $\gets$ max(\textit{$O_{CM}$}, \textit{q.length})\;
					\textit{$CM$} $\gets$ max(\textit{$CM$}, $w_1 \times q.length + w_2 \times q.SLO$)\;
				}
				\Else{
					\textit{batches.append($batch_c$)}\;
					\textit{$batch_c$} $\gets$ \{\textit{q.index}\}\;
					\textit{$L_{CM}$} $\gets$ \textit{q.SLO}\;
					\textit{$O_{CM}$} $\gets$ \textit{q.length}\;
					\textit{$CM$} $\gets$ $w_1 \times q.length + w_2 \times q.SLO$\;
				}
				Dynamically adjust $batch\_size$ according to the value of $CM$\;
			}
			
			\If{\textit{$batch_c \neq \emptyset $} }{
				\textit{batches.append($batch_c$)}\;
			}
			
			\Return{\textit{batches}}\;
		}
        
	\end{algorithm}
	
 	\vspace{-0.6cm}
	Based on the SLO-ODBS algorithm, different scheduling objectives can be addressed by adjusting the values of $w_1$ and $w_2$. Specifically, when $w_1=0$, we developed the SLO Dynamic Batch Schedule (SLO-DBS) algorithm to reduce the SLO violation rate by efficiently arranging inference requests. Conversely, when $w_2=0$, we designed the Output-Driven Dynamic Batch Schedule (ODBS) algorithm to minimize inference latency by skillfully merging requests based on the predicted output length.
	
	\vspace{-0.4cm}
	\subsection{LLM Deployer} \label{deploy}
	
	The $LLM$ $deployer$ mainly starts during the LLM deployment phase. It arranges the layers of the LLM according to the topology of the current system's hardware accelerators and the computational characteristics of each LLM layer. This process establishes a suitable device mapping to allocate each LLM layer to appropriate hardware accelerators, thereby achieving the objectives of enhancing hardware utilization and reducing latency. Specifically, it can be defined as follows: given a hardware network represented by a graph $G = (D, E)$, where $D = \{d_1, d_2, \ldots, d_n\}$ is a set of hardware devices, and $E$ is the set of edges connecting the hardware. The large model $llm_i$ has memory requirements $M(llm_i)$ and the number of required layers $Layer(llm_i)$. Each node $d_i$ possesses the following attributes: $Memory(d_i)$ denotes the available memory at node $d_i$, and $performance(d_i)$ represents the computational capacity of node $d_i$. The objective is to find a device allocation scheme $S\subseteq D$ that minimizes processing time while satisfying memory constraint: 
	
	\begin{equation}
		\sum_{i=1}^{|S|}\text{Memory}(d_i) \geq M.
	\end{equation}

	It is straightforward to see that this problem is a dynamic programming problem. Based on this, we designed the HELR as shown in Algorithm \ref{helr}. This algorithm is capable of ascertaining the most effective configuration for the deployment of an LLM, considering the current cluster node topology and the specific LLM intended for deployment. The algorithm is divided into three parts: \textbf{1)} the initialization phase (lines 1-2), \textbf{2)} the dynamic programming phase (lines 3-15), and \textbf{3)} the device map update phase (lines 16-19).
	

	In stage \textbf{1)}, some necessary parameters are set and obtain the current model information and cluster topology structure.
	
	In stage \textbf{2)}, a two-dimensional array $dp$ is maintained, where $dp[mark][i]$ denotes the Minimal latency from the initial state to the device node $d_i$ and $Performance(d)$ denotes processing performance of device $d$. The dynamic programming recurrence relation can be expressed as:
	\begin{equation} \label{dp}
		\mathrm{dp}[mark][i]=\min_{1 \leq j\leq |S|}\left(dp[mark][i], dp[i][j] + Latency(E[i][j]) + p \times \frac{layers[i]\times m}{performance(i)}\right).
	\end{equation}
	Therefore, our goal is to minimize latency:
	\begin{equation}
		\min_{1\leq i \leq |S|}\left(dp[2^{k}-1][i] + \sum_{j=1}^{|S|}\left(Latency(E[i][j])+p \times \frac{layers[i]\times m}{performance(i)}\right)\right),
	\end{equation}
	where $layers[i]$ represents the number of layers assigned to device node $d_i$, $Latency(E[i][j])$ represents the communication delay between device node $d_i$ and  $d_j$, and $p$ adjusts processing performance-time relationship. The algorithm updates the array $dp$ and the current state using Eq. (\ref{dp}).

	In stage \textbf{3)}, the optimal allocation state $best\_state$ is recorded to ensure the best possible utilization of resources. The $Device\_map$ is then updated using the information from $layers$ and $S$. This update process ensures that the deployment configuration reflects the optimal state, leading to improved efficiency and performance in the subsequent stages.

	Similar to the batch processing algorithm, different deployment objectives can be achieved by adjusting the values of $a_1$ and $a_2$ in the HELR algorithm. Due to the sequential nature of inference, deploying on the minimum number of GPUs possible can effectively improve GPU utilization. Therefore, to optimize GPU utilization, setting $a_1$ to 0 while updating the $dp$ array can significantly enhance utilization. This configuration of $a_1$ forms the High-Efficiency Resource Allocation (HE) algorithm, which is suitable for environments with limited resources. Conversely, to fulfill the minimum latency requirement without considering expenses, the weight of $a_1$ can be increased. For example, setting $a_1$ to 10:1 establishes the Low-Latency Resource Allocation (LR) algorithm, which prioritizes latency by assigning a high weight to $a_1$.
	\vspace{-0.4cm}
	\begin{algorithm}[htbp]
		\caption{High-Efficiency Low-Latency Resource Allocation Algorithm (HELR)}
		\label{helr}
		\KwIn{
			$M$: Memory requirement of LLM, $Layer(M)$: Number of layers in the large model $M$\\
			$G(D,E)$: Graph representing the hardware platform, $E$: Connections between various nodes\\
			$D$: Hardware device nodes, $Latency(E[i][j])$: Communication latency between node $i$ and node $j$\\
			$Performance(d)$: Processing performance of device $d$, $Memory(d)$: Available memory of device $d$\\
		}
		\KwOut{$Device\_map$: Mapping of layers to devices}
		
		\SetKwFunction{HELR}{HELR}
		\SetKwProg{Fn}{Procedure}{:}{}
		\Fn{\HELR{$M, Layer(M), G(D,E), E, D, Latency(E[i][j]), Performance(d), Memory(d)$}}{
			Initialize $best\_state \leftarrow \infty$, Initialize $Device\_map \leftarrow \emptyset$ \;
			
			\For{each $n$ from 1 to $|D|$}{
				$S_n$ is the subset of $D$ with size $n$ \; 
				\If{the total memory of nodes in $S_n$ is more than $M$}{skip to the next subset \;} 
				Initialize $dp[mark][i] \gets \infty$ \; 
				\For{each mark from $1$ to $2^n-1$}{
					Sort the nodes in $S_n$ in descending order by performance and memory\;
					Calculate the memory per layer $m \leftarrow \frac{M}{Layer(M)}$\;
					\For{each $i$ from 1 to  $|S_n|$}{
						\For{each $j$ from 1 to $|S_n|$ where $j \neq i$}{
							
							\tcp{ $T$ is the memory reserved for KV Cache}
							Calculate the maximum layers assignable to node $i$: $layers[i] \leftarrow \min(Layer(M), \frac{Memory(i) - T}{m})$\;
							
							Calculate the latency $l$ using the formula: $l = dp[i][j] + Latency(E[i][j]) + p \times \frac{layers[i] \times m}{Performance(i)}$\;
							Update $dp[mark][i] \leftarrow \min(dp[mark][i], l)$\;
						}
					}
				}
				
				$current\_state \leftarrow \min(dp[2^n-1][i] + \sum_{j = 1}^{|S_n|}(Latency(E[i][j]) + p \times \frac{layers[i] \times m}{Performance(i)})) $\;
				
				\If{$current\_state < best\_state$}{
					$best\_state \leftarrow current\_state$\;
					Update $Device\_map$ with $layers$ and nodes in $S_n$\;
				}
			}
			\Return{\textit{Device\_map}}\;
		}
	\end{algorithm}

	\vspace{-0.8cm}
	\section{Implementation and Evaluations} \label{eva}
	\subsection{Evaluation Setup}
	\hspace*{1.1em} \textbf{Testbed.} We use a local cluster (consisting of 4 Nvidia RTX 3090 GPUs) as the test platform for conducting extensive experiments. To differentiate GPU performance in our experiments, we set different performance limits for the GPUs, as detailed in the Table \ref{tab:cluster}.

	\textbf{ML models.} For the LLM inference service, we select the ChatGLM2-6B \cite{du2022glm}, which is currently stable and widely recognized, to perform inference tasks. As the ChatGLM2-6B model lacked batch inference capabilities at the time, we make adjustments to its inference code to enable batch processing for future experiments.
	
	\textbf{SLO.} In our experiments, we define the SLO as the requirement for an inference request to receive a complete answer within a certain time frame. To better approximate real-world scenarios, we designed different SLOs for different inference requests, ranging from 1 second to 350 seconds, ensuring that each inference request's SLO is completely random.
	
	\textbf{Resource monitoring.} Furthermore, it is essential to maintain continuous real-time monitoring of individual GPUs within the cluster, focusing on metrics such as GPU utilization, GPU memory usage, and the execution time of inference programs. To achieve this, we utilized Nvidia's provided interfaces to develop a script capable of real-time monitoring of GPU information across the cluster.
	
         \vspace{-0.3cm}
	\begin{table}[htbp]
		\scriptsize
		\centering
		\begin{tabular}{>{\centering\arraybackslash}p{2.2cm} >{\centering\arraybackslash}p{2.3cm} >{\centering\arraybackslash}p{2.3cm} >{\centering\arraybackslash}p{2.3cm} >{\centering\arraybackslash}p{2.3cm} >{\centering\arraybackslash}p{3cm}}
			\rowcolor{customblue!20} 
			\toprule
			& \textbf{GPU\#0} & \textbf{GPU\#1} & \textbf{GPU\#2} & \textbf{GPU\#3} & \textbf{Maximum power} \\
			\midrule
			\rowcolor{customblue!10} 
			\textbf{GPU\#0} & X & PIX & NODE & NODE & 350 W \\ 
			\rowcolor{customblue!20}
			\textbf{GPU\#1} & PIX & X & NODE & NODE & 300 W \\ 
			\rowcolor{customblue!10}
			\textbf{GPU\#2} & NODE & NODE & X & PIX & 250 W \\ 
			\rowcolor{customblue!20}
			\textbf{GPU\#3} & NODE & NODE & PIX & X & 150 W \\ 
			\bottomrule
		\end{tabular}
		\vspace{0.3cm}
		\caption{Cluster network topology. There are three different types of connections: 1) X  = Self, 2) PIX  = Connection traversing at most a single PCIe bridge, and 3) NODE = Connection traversing PCIe as well as the interconnect between PCIe Host Bridges within a NUMA node.}
		\label{tab:cluster}
	\end{table}
	\vspace{-1.5cm}
	\subsection{Baselines and Metrics}
	\hspace*{1.0em} \textbf{Baselines.} We construct three versions of the UELLM prototype on a local cluster: 1) UELLM-deploy (UD), which only uses the HELR model deployment algorithm. 2) UELLM-batch (UB), which uses the SLO-ODBS batching algorithm. 3) UELLM-all (UA), which employs both the HELR model deployment algorithm and the SLO-ODBS batching algorithm. Currently, there are almost no systems similar to ours, so we chose the current SOTA method Morphling (Mor)\cite{morphling} and the batching algorithm in $S^3$\cite{jin2023s3} as our baselines for comparison with UD, UB, and UA. 
	
	\textbf{Metrics.} We adopt four widely used metrics to evaluate performance: 1) \textbf{Latency}, denoting the time taken for the system to respond to a request, encompassing the duration from request initiation to the commencement of processing and result delivery. Lower latency enhances the user experience of LLM inference services. 2) \textbf{Throughput}, quantified by the number of tokens processed per second. Higher throughput means the system capacity to manage requests or data within a specific timeframe, thereby improving overall processing efficiency. 3) \textbf{GPU utilization}, where high utilization typically indicates efficient execution of computational tasks, while low utilization may suggest underutilized resources or bottlenecks necessitating further optimization. 4) \textbf{SLO violation}, a crucial metric for assessing user experience. A lower default rate generally signifies a stable and dependable service, leading to higher user satisfaction. Conversely, a high default rate can result in diminished user experience or even user attrition.
	
	\vspace{-0.2cm}
	\subsection{Experiment Analyses}
\begin{figure}[htbp]
		\centering
		\begin{subfigure}[b]{0.24\textwidth}
			\centering
			\includegraphics[width=\textwidth]{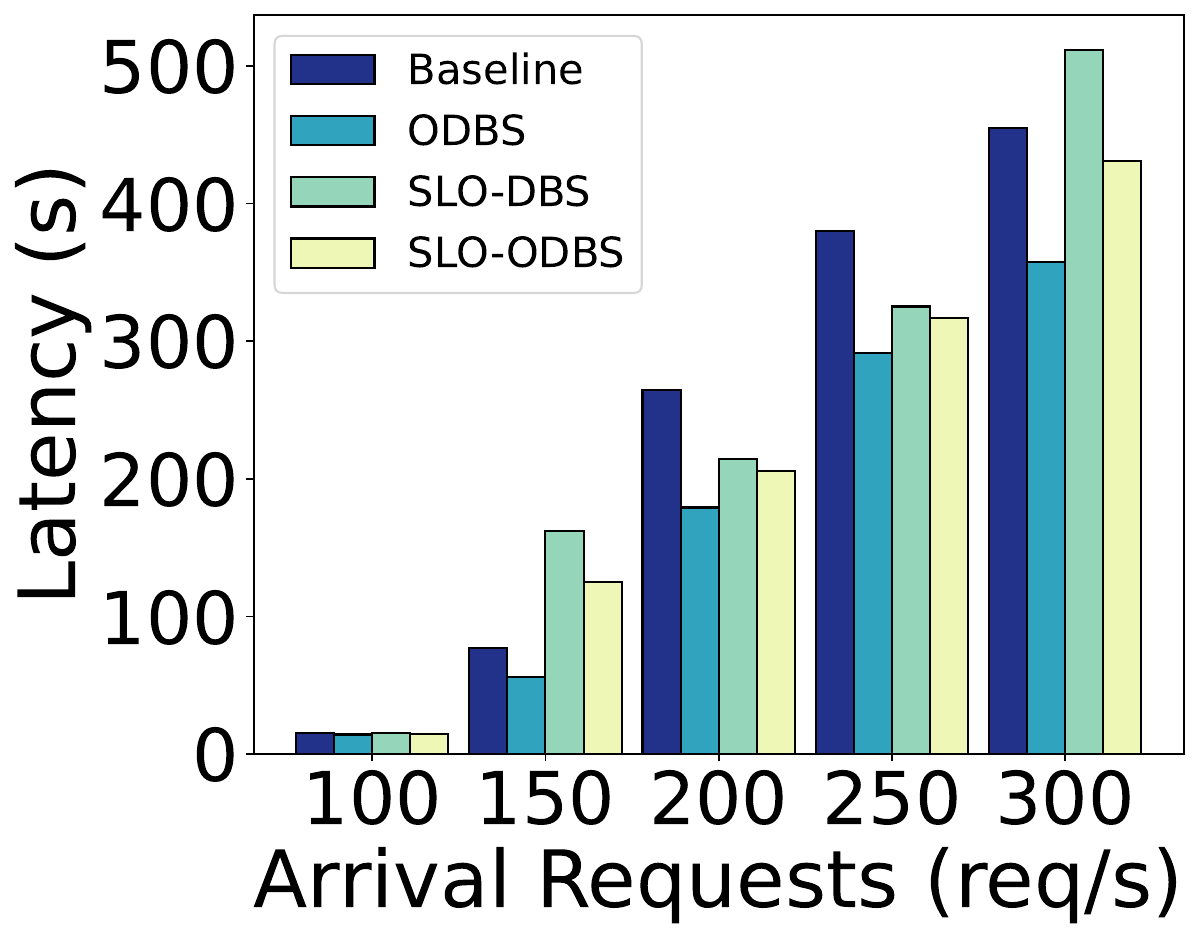}
			\caption{Latency}
			\label{fig:arlatency}
		\end{subfigure}
		\hfill
		\begin{subfigure}[b]{0.23\textwidth}
			\centering
			\includegraphics[width=\textwidth]{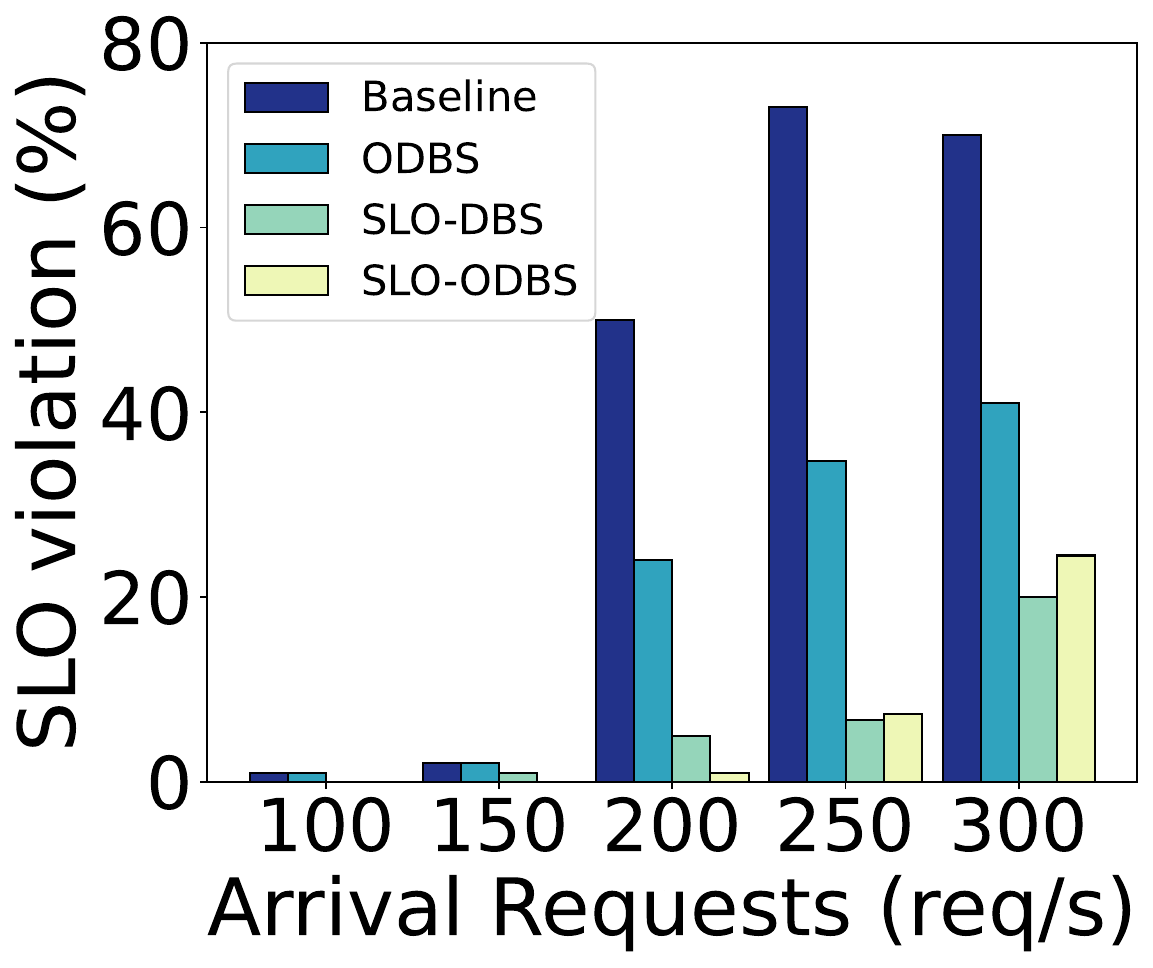}
			\caption{SLO violation}
			\label{fig:ar slo}
		\end{subfigure}
		\hfill
		\begin{subfigure}[b]{0.24\textwidth}
			\centering
			\includegraphics[width=\textwidth]{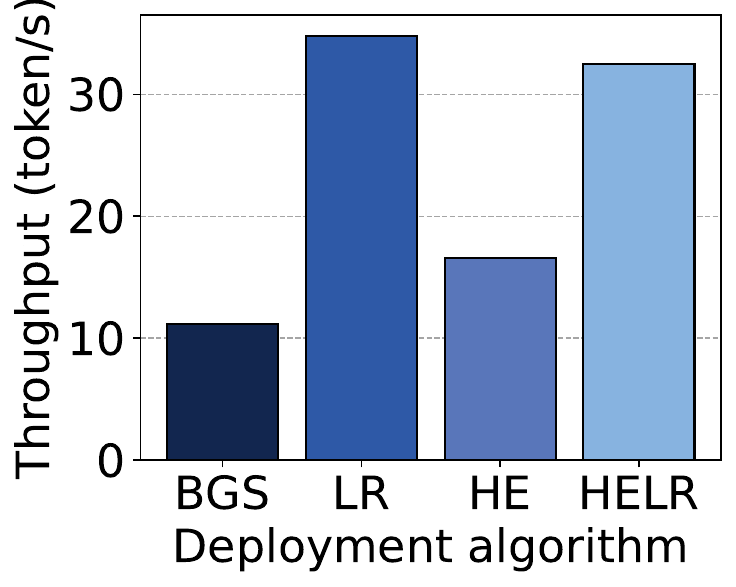}
			\caption{Throughput}
			\label{fig:he throughput}
		\end{subfigure}
		\hfill
		\begin{subfigure}[b]{0.24\textwidth}
			\centering
			\includegraphics[width=\textwidth]{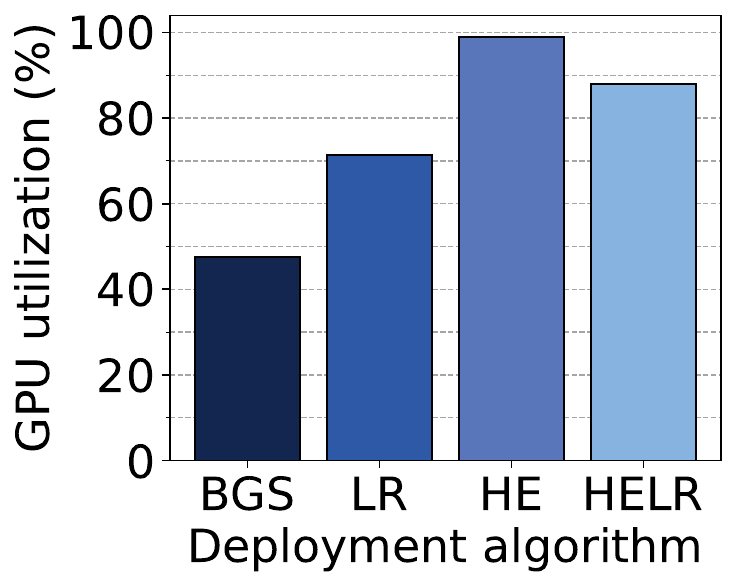}
			\caption{GPU utilization}
			\label{fig:he gpu}
		\end{subfigure}
		\caption{Performance comparison of different batching algorithms and deployment algorithms. (a) Latency, (b) SLO violation under various batching algorithms, while (c) Throughput, and (d) GPU utilization under different deployment algorithms.}
		\label{fig:ar}
		\vspace{-0.5cm}
	\end{figure}

\begin{figure}[htbp]
		\centering
		\begin{subfigure}[b]{0.24\textwidth}
			\centering
			\includegraphics[width=\textwidth]{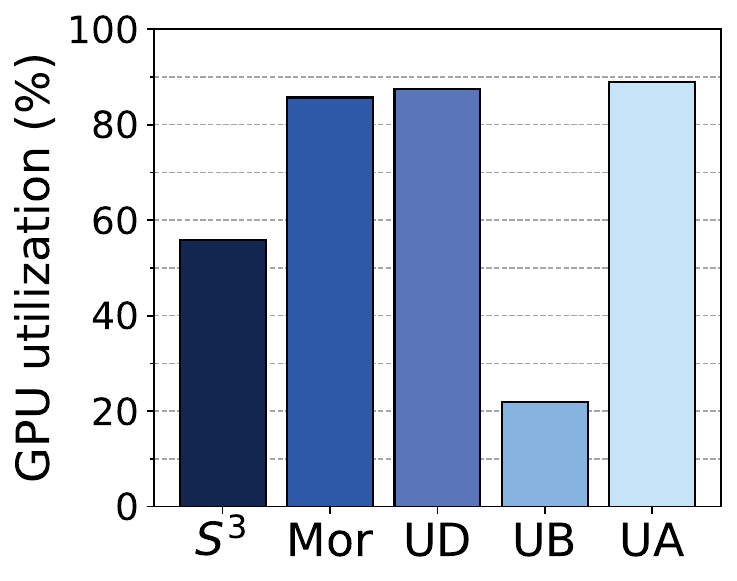}
			\caption{GPU utilization}
			\label{fig:7GPU Utilization}
		\end{subfigure}
		\hfill
		\begin{subfigure}[b]{0.24\textwidth}
			\centering
			\includegraphics[width=\textwidth]{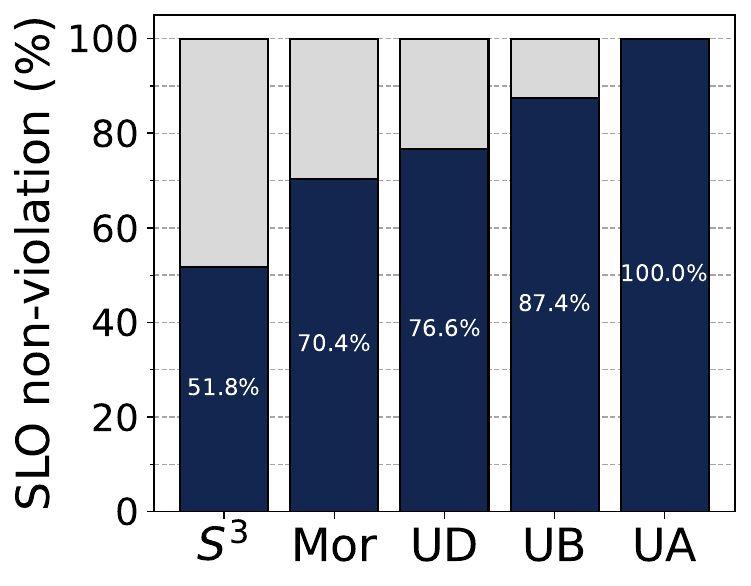}
			\caption{SLO Violation}
			\label{fig:7slo}
		\end{subfigure}
		\hfill
		\begin{subfigure}[b]{0.24\textwidth}
			\centering
			\includegraphics[width=\textwidth]{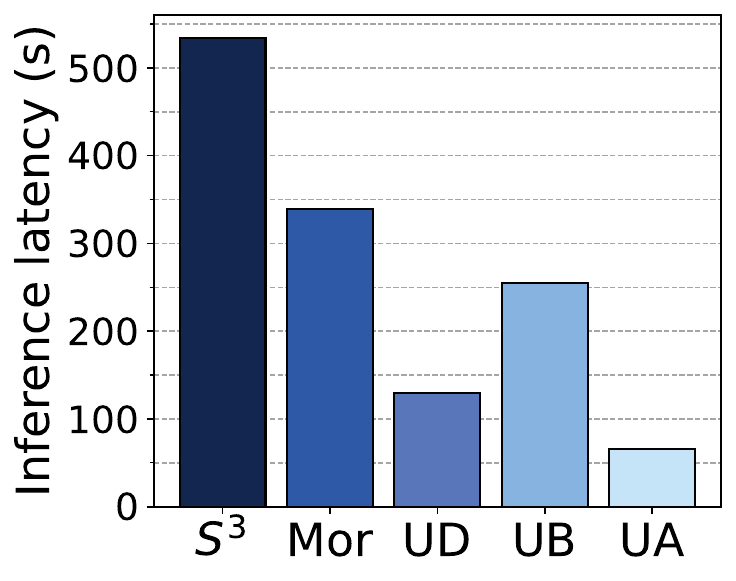}
			\caption{Latency}
			\label{fig:7latency}
		\end{subfigure}
		\hfill
		\begin{subfigure}[b]{0.24\textwidth}
			\centering
			\includegraphics[width=\textwidth]{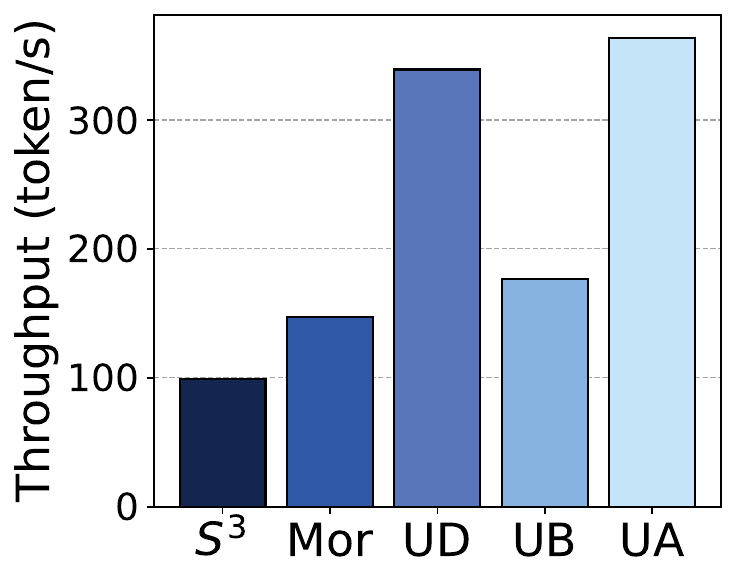}
			\caption{Throughput}
			\label{fig:7throughput}
		\end{subfigure}
		\caption{Comparison results of various metrics between \(S^3\), Morphling, UD, UB, and UA. Metrics include (a) GPU utilization, (b) SLO non-violation, (c) inference latency, and (d) throughput.}
		\label{fig:7}
		\vspace{-0.5cm}
	\end{figure}

	To minimize randomness, each experiment was repeated 5 times, and each data point in Fig. \ref{fig:ar} and Fig. \ref{fig:7} represents the average of these 5 trials. 
	Before conducting the main experiments, we validated the effectiveness and advancement of the batching algorithms and LLM deployment algorithms by two comparisons: \textbf{1)} We compared SLO-ODBS, SLO-DBS, and ODBS with the default batching algorithm (FIFO) on scheduling metrics: latency and SLO violation rate. \textbf{2)} We compared LR, HE, and HELR with the baseline deployment algorithm, Greedy Scheduling Algorithm (BGS), on deployment metrics: throughput and GPU utilization. Fig. \ref{fig:arlatency} and Fig. \ref{fig:ar slo} show that under high request loads, by reasonably combining requests, SLO-ODBS reduces the number of iterations and memory overhead, maintaining low latency similar to the ODBS. At the same time, SLO-ODBS schedules requests according to SLOs, achieving a low SLO violation rate close to the SLO-DBS. Because HELR selects more reasonable resource allocation and deployment methods, Fig. \ref{fig:he throughput} and Fig. \ref{fig:he gpu} show that HELR maintains utilization close to HE while achieving throughput similar to the LR.
	\vspace{-0.3cm}
	
	\vspace{-0.4cm}
	Fig. \ref{fig:7} compares the differences between UELLM and the SOTA algorithms $S^3$ and Morphling across various metrics. Fig. \ref{fig:7GPU Utilization} illustrates the comparison of GPU utilization, showing the average results of the algorithms over five different time periods. It is observed that due to $S^3$ and UB focusing solely on batch scheduling, their GPU utilization is significantly lower compared to other baselines, indicating inefficiency in resource utilization. Morphling uses meta-learning to search for the optimal configuration and conducts stress testing, while UA and UD use the HELR algorithm to select the best resource configuration based on the current node topology and model characteristics. Therefore, the results of these three are very close. Overall, UELLM can achieve deployment effects similar to Morphling with lower costs.
	\vspace{-0.2cm}
	
	Fig. \ref{fig:7slo} describes the comparison of SLO violation rates. UB, using only the SLO-ODBS batching algorithm with the default deployment algorithm, can meet the SLO for 87.4\% of requests, serving as a baseline for analyzing the performance of the other four algorithms. Compared to UB, $S^3$ only considers memory optimization without considering the SLO of each request and deployment strategy, resulting in the poorest performance. The results of Morphling and UD are almost identical because both only consider deployment strategies without considering the SLOs of different requests. Although their inference delays are lower in Figure 5d, they still have a higher SLO violation rate compared to UB. UA represents the optimal state of UELLM, considering both SLO and deployment strategy. Thus, in five experiments, UA meets the SLO for all requests, achieving the best performance. Overall, compared to $S^3$ and Morphling, UELLM optimizes the SLO violation rate by 29.6\% to 48.2\%.

	Fig. \ref{fig:7latency} and Fig. \ref{fig:7throughput} describe the comparison of inference latency and throughput, both metrics indicating the inference speed of the system, and are thus discussed together. We use $S^3$ and Morphling as benchmarks to analyze the performance of the other three algorithms. Compared to Morphling, UD reduces stress testing, which in turn reduces stress testing time, leading to lower overall inference latency and improved throughput. Compared to $S^3$, UD optimizes resource allocation, improving utilization and communication latency, significantly reducing inference latency. Compared to Morphling and $S^3$, UB uses the SLO-ODBS algorithm to optimize batch combinations, reducing iterations, lowering inference latency, and improving throughput. UA shows the best performance because it reduces latency and improves throughput from both reasonable batch combination and resource allocation dimensions. UA uses the HELR algorithm to optimize resource allocation and the SLO-ODBS algorithm to optimize batch combinations, achieving the best results in five experiments. Overall, compared to $S^3$ and Morphling, UELLM reduces the inference latency by 72.3\% to 90.3\% and improves throughput by $1.92\times$ to $4.98\times$.
         \vspace{-0.2cm}
	\section{Conclusion}
	In this paper, we propose UELLM, a framework that integrates request batching and LLM deployment. UELLM is designed to maximize throughput, reduce inference latency, lower SLO violation rates, and minimize memory wastage. We introduce the HELR LLM deployment algorithm and the SLO-ODBS batching algorithm. The SLO-ODBS algorithm optimizes batch composition, while HELR optimizes resource utilization during deployment, ensuring UELLM maintains high-quality service in terms of inference latency, throughput, and SLO violation rates. Our experiments demonstrate that UELLM outperforms the state-of-the-art in efficiently utilizing resources and reducing SLO violation rates. This approach has the potential to significantly enhance the efficiency and reliability of LLM-based inference services in cloud computing environments.\\
	\\
	\indent\textbf{Software Availability:} The codes have been open-sourced to \url{https://github.com/HYIUYOU/UELLM} for research usage.
	
	\section{Acknowledgments} This work is supported by National Key R \& D Program of China (No. 2021YFB3300200), the National Natural Science Foundation of China (No. 62072451, 62102408, 92267105), and Guangdong Basic and Applied Basic Research Foundation (No. 2024A1515010251, 2023B1515130002).
	\printbibliography

@inproceedings{tabi,
author = {Wang, Yiding and Chen, Kai and Tan, Haisheng and Guo, Kun},
title = {Tabi: An Efficient Multi-Level Inference System for Large Language Models},
year = {2023},
isbn = {9781450394871},
url = {https://doi.org/10.1145/3552326.3587438},
doi = {10.1145/3552326.3587438},
booktitle = {Proceedings of the Eighteenth European Conference on Computer Systems (EuroSys 23)},
pages = {233–248},
numpages = {16}
}

@inproceedings {cocktail,
author = {Jashwant Raj Gunasekaran and Cyan Subhra Mishra and Prashanth Thinakaran and Bikash Sharma and Mahmut Taylan Kandemir and Chita R. Das},
title = {Cocktail: A Multidimensional Optimization for Model Serving in Cloud},
booktitle = {19th USENIX Symposium on Networked Systems Design and Implementation (NSDI 22)},
year = {2022},
isbn = {978-1-939133-27-4},
pages = {1041--1057},
url = {https://www.usenix.org/conference/nsdi22/presentation/gunasekaran},
month = apr
}

@inproceedings{wonik,
  title={SLO-aware inference scheduler for heterogeneous processors in edge platforms},
  author={Seo, Wonik and Cha, Sanghoon and Kim, Yeonjae and Huh, Jaehyuk and Park, Jongse},
  booktitle={ACM Transactions on Architecture and Code Optimization (TACO 21)},
  volume={18},
  number={4},
  pages={1--26},
  year={2021},
  
}

@inproceedings{batch,
  title={Batch: Machine learning inference serving on serverless platforms with adaptive batching},
  author={Ali, Ahsan and Pinciroli, Riccardo and Yan, Feng and Smirni, Evgenia},
  booktitle={International Conference for High Performance Computing, Networking, Storage and Analysis (SC 20)},
  pages={1--15},
  year={2020},
  organization={IEEE}
}

@inproceedings {mark,
author = {Chengliang Zhang and Minchen Yu and Wei Wang and Feng Yan},
title = {{MArk}: Exploiting Cloud Services for {Cost-Effective}, {SLO-Aware} Machine Learning Inference Serving},
booktitle = {2019 USENIX Annual Technical Conference (USENIX ATC 19)},
year = {2019},
isbn = {978-1-939133-03-8},
pages = {1049--1062},
url = {https://www.usenix.org/conference/atc19/presentation/zhang-chengliang},
month = jul
}

@inproceedings{morphling,
author = {Wang, Luping and Yang, Lingyun and Yu, Yinghao and Wang, Wei and Li, Bo and Sun, Xianchao and He, Jian and Zhang, Liping},
title = {Morphling: Fast, Near-Optimal Auto-Configuration for Cloud-Native Model Serving},
year = {2021},
isbn = {9781450386388},
url = {https://doi.org/10.1145/3472883.3486987},
doi = {10.1145/3472883.3486987},
booktitle = {Proceedings of the ACM Symposium on Cloud Computing (SoCC 21)},
pages = {639–653},
numpages = {15},

}

@inproceedings {choi,
author = {Seungbeom Choi and Sunho Lee and Yeonjae Kim and Jongse Park and Youngjin Kwon and Jaehyuk Huh},
title = {Serving Heterogeneous Machine Learning Models on {Multi-GPU} Servers with {Spatio-Temporal} Sharing},
booktitle = {2022 USENIX Annual Technical Conference (USENIX ATC 22)},
year = {2022},
isbn = {978-1-939133-29-53},
pages = {199--216},
url = {https://www.usenix.org/conference/atc22/presentation/choi-seungbeom},
month = jul
}

@inproceedings{du2022glm,
    title = "{GLM}: General Language Model Pretraining with Autoregressive Blank Infilling",
    author = "Du, Zhengxiao  and
      Qian, Yujie  and
      Liu, Xiao  and
      Ding, Ming  and
      Qiu, Jiezhong  and
      Yang, Zhilin  and
      Tang, Jie",
    booktitle = "Proceedings of the 60th Annual Meeting of the Association for Computational Linguistics (ACL 22)",
    month = may,
    year = "2022",
    url = "https://aclanthology.org/2022.acl-long.26",
    doi = "10.18653/v1/2022.acl-long.26",
    pages = "320--335",
}

@inproceedings{gpt,
 author = {Brown, Tom and Mann, Benjamin and Ryder, Nick and Subbiah, Melanie and Kaplan, Jared D and Dhariwal, Prafulla and Neelakantan, Arvind and Shyam, Pranav and Sastry, Girish and Askell, Amanda and Agarwal, Sandhini and Herbert-Voss, Ariel and Krueger, Gretchen and Henighan, Tom and Child, Rewon and Ramesh, Aditya and Ziegler, Daniel and Wu, Jeffrey and Winter, Clemens and Hesse, Chris and Chen, Mark and Sigler, Eric and Litwin, Mateusz and Gray, Scott and Chess, Benjamin and Clark, Jack and Berner, Christopher and McCandlish, Sam and Radford, Alec and Sutskever, Ilya and Amodei, Dario},
 booktitle = {Advances in Neural Information Processing Systems 33 (NeurlPS 20)},
 pages = {1877--1901},
 title = {Language Models are Few-Shot Learners},
 url = {https://proceedings.neurips.cc/paper_files/paper/2020/file/1457c0d6bfcb4967418bfb8ac142f64a-Paper.pdf},
 volume = {33},
 year = {2020}
}

@inproceedings{vaswani2023attention,
  title={Attention is all you need},
  author={Vaswani, Ashish and Shazeer, Noam and Parmar, Niki and Uszkoreit, Jakob and Jones, Llion and Gomez, Aidan N and Kaiser, {\L}ukasz and Polosukhin, Illia},
  booktitle={Advances in neural information processing systems (NeurlPS 17)},
  volume={30},
  year={2017}
}

@inproceedings{jin2023s3,
  title={$ S^3 $: Increasing GPU Utilization during Generative Inference for Higher Throughput},
  author={Jin, Yunho and Wu, Chun-Feng and Brooks, David and Wei, Gu-Yeon},
  booktitle={Advances in Neural Information Processing Systems (NeurlPS 23)},
  volume={36},
  pages={18015--18027},
  year={2023}
}

@inproceedings {clipper,
author = {Daniel Crankshaw and Xin Wang and Guilio Zhou and Michael J. Franklin and Joseph E. Gonzalez and Ion Stoica},
title = {Clipper: A {Low-Latency} Online Prediction Serving System},
booktitle = {14th USENIX Symposium on Networked Systems Design and Implementation (NSDI 17)},
year = {2017},
isbn = {978-1-931971-37-9},
pages = {613--627},
url = {https://www.usenix.org/conference/nsdi17/technical-sessions/presentation/crankshaw},
month = mar
}

@inproceedings{vllm,
  title={Efficient memory management for large language model serving with pagedattention},
  author={Kwon, Woosuk and Li, Zhuohan and Zhuang, Siyuan and Sheng, Ying and Zheng, Lianmin and Yu, Cody Hao and Gonzalez, Joseph and Zhang, Hao and Stoica, Ion},
  booktitle={Proceedings of the 29th Symposium on Operating Systems Principles (SOSP 23)},
  pages={611--626},
  year={2023}
}

@inproceedings{metaai,
  title={Sustainable ai: Environmental implications, challenges and opportunities},
  author={Wu, Carole-Jean and Raghavendra, Ramya and Gupta, Udit and Acun, Bilge and Ardalani, Newsha and Maeng, Kiwan and Chang, Gloria and Aga, Fiona and Huang, Jinshi and Bai, Charles and others},
  booktitle ={Proceedings of Machine Learning and Systems (MLSys 22)},
  volume={4},
  pages={795--813},
  year={2022}
}

@inproceedings{sheng2023flexgen,
  title={Flexgen: High-throughput generative inference of large language models with a single gpu},
  author={Sheng, Ying and Zheng, Lianmin and Yuan, Binhang and Li, Zhuohan and Ryabinin, Max and Chen, Beidi and Liang, Percy and R{\'e}, Christopher and Stoica, Ion and Zhang, Ce},
  booktitle={International Conference on Machine Learning (ICML 23)},
  pages={31094--31116},
  year={2023},
  organization={PMLR}
}
	
\end{document}